\documentclass[prl,twocolumn,showpacs,superscriptaddress]{revtex4}
\usepackage{graphicx,color}
\usepackage{amsmath}
\usepackage{amssymb}

\usepackage{hyperref}

\begin{document}

\title{Spin based heat engine: demonstration of multiple rounds of algorithmic cooling.}

\author{C.A. Ryan }
\affiliation{Institute for Quantum Computing and Dept. of Physics, University of Waterloo, Waterloo, ON, N2L 3G1, Canada.}

\author{O. Moussa}
\affiliation{Institute for Quantum Computing and Dept. of Physics, University of Waterloo, Waterloo, ON, N2L 3G1, Canada.}

\author{J. Baugh}
\affiliation{Institute for Quantum Computing and Dept. of Physics, University of Waterloo, Waterloo, ON, N2L 3G1, Canada.}

\author{R. Laflamme}
\affiliation{Institute for Quantum Computing and Dept. of Physics, University of Waterloo, Waterloo, ON, N2L 3G1, Canada.}
\affiliation{Perimeter Institute for Theoretical Physics, Waterloo, ON, N2J 2W9, Canada}

\date{\today}

\begin{abstract}
We show experimental results demonstrating multiple rounds of heat-bath algorithmic cooling in a 3 qubit solid-state nuclear magnetic resonance quantum information processor.  By dynamically pumping entropy out of the system of interest and into the heat-bath, we are able show purification of a single qubit to a polarization 1.69 times that of the heat-bath and thus go beyond the Shannon bound for closed system cooling.  The cooling algorithm implemented requires both high fidelity coherent control and a  deliberate controlled interaction with the environment.  We discuss the improvements in control that allowed this demonstration.  This experimental work shows that given this level of  quantum control in systems with sufficiently large polarizations, nearly pure qubits should be achievable.     
\end{abstract}


\maketitle

Using quantum mechanics to process information promises the possibility to dramatically speed-up certain computations and simulations \cite{IkeandMike}.  Many experimental paths are being pursued in the goal of coherently manipulating quantum systems \cite{QIP:2004a}.   The standard circuit based model has certain experimental criteria \cite{divincenzo:2000a}: one of which is the ability to initialize pure fiducial quantum states.  This is needed not only to create the initial state for many quantum algorithms, but it is also necessary to have pure qubits on demand throughout the computation in order to compute fault-tolerantly in the presence of errors \cite{Preskill:1998a}.  However, many physical implementations are able to initialize only mixed states with a certain bias towards the desired state.  In these cases it will almost certainly be necessary to run some protocol to purify the qubits.  Aside from quantum information purposes, the ability to increase the bias of nuclear spins is fundamentally important in nuclear magnetic resonance (NMR) where small signal to noise ratios are usually overcome with signal averaging.   A boost in the initial bias by a factor $b$ would reduce the experiment time by $b^2$.

A potential solution is algorithmic, whereby a series of logic gates can lower the temperature of a subset of the qubits.  If the relaxation rate back to thermal equilibrium ($T_1$) is sufficiently longer than the cooling operations, then it is possible to cool a subset far below their thermal equilibrium bias.   This algorithmic cooling is essentially classical and is based on early work from von Neumann \cite{vonNeumann:1956a}.  If the bits start with some bias $\epsilon$, so the probability of being in the state 0 or spin up is $P_\uparrow = \frac{1+\epsilon}{2}$ and $P_\downarrow = \frac{1-\epsilon}{2}$, then the application of a logic gate can compress the uncertainty into some fraction of the qubits and increase the bias on the rest.  Using these ideas it was shown that by starting with a sufficient number of qubits, it is possible to initialize a small number of qubits to a fiducial state with near certainty \cite{Schulman:1998a,Schulman:1999a}.  However, for the starting biases typical of room temperature NMR,  that sufficient number is an impractically large number e.g., to purify only one qubit requires $\approx 10^{12}$ spins.  In a closed system, the compression step is limited by the Shannon bound: the total entropy of the system is conserved.  The level of purification is usually further limited if the compression step is restricted to unitary transformations \cite{Sorensen:1989a}.  As a relevant example, with three qubits, each starting with the same polarization $\epsilon$, it is not possible to amplify the bias of one qubit to more than $1.5\epsilon$.

If, on the other hand, we consider an open system, and allow the ability to pump entropy out of the system in a controlled manner, then we can surpass the Shannon bound.  Suppose that some of the qubits have a fast relaxation time and quickly return to equilibrium with the heat-bath. Every compression step cools some subset of qubits and heats up the remainder of the qubits above the heat-bath temperature.  If these heated qubits are then allowed to relax back to the heat-bath temperature, the total entropy of the qubit system has decreased.   The cooling algorithm then consists of alternating rounds of cooling and compression \cite{Boykin03192002,Fernandez:2004a,Rempp:2007a,Kaye:2007a}.   Recently Schulman et al.  \cite{schulman:120501} have shown an optimal algorithm, the partner pairing algorithm (PPA), for the scenario of having one special purpose reset qubit.  This optimality also allowed them to show a crucially important threshold effect: given $n$ qubits and a heat-bath bias of $\epsilon \gg 2^{-n}$ then it is possible to almost perfectly purify the system with resources growing polynomially in $n$; whereas, if $\epsilon \ll 2^{-n}$, the maximum bias achievable on one qubit is $\epsilon2^{n-2}$ \cite{moussa:2005a}.  A similar system with differential relaxation rates has been considered for error correction purposes and could also be used for purification \cite{Sarovar:2005ys}.    Several parts of the cooling algorithms, including both the compression step \cite{Chang:2001a} and the reset \cite{Brassard:2005a}, have been experimentally demonstrated using NMR quantum information processors (QIP).    These were combined by Baugh et al. \cite{Baugh:2005a} to show one round of cooling three qubits and the compression step.   However, sufficient control was lacking to demonstrate multiple rounds of cooling and compression.  Here, we present experimental results showing multiple rounds of resetting and compression steps allowing us to go beyond the Shannon bound for the first time.  
 
NMR offers one of the most advanced implementations of a QIP with high fidelity control and several qubits \cite{vandersypen:1037}.  The qubits are nuclear spins in a bulk ensemble sample where many, ideally identical, copies of the processor are manipulated in parallel.  Readout consists of measuring the expectation value of operators averaged over the sample.   A large static magnetic field provides the quantization axis and for spin 1/2 nuclei, two Zeeman energy levels.  In thermal equilibrium at room termpature there is a very small bias or polarization which the algorithmic cooling circuit can amplify.  

\begin{figure}[htbp]
\includegraphics[scale=0.5]{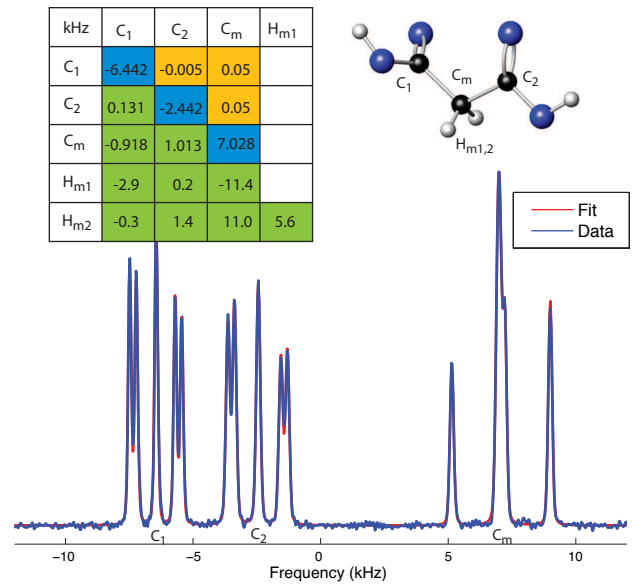}
\caption{\label{RefSpectrum}
The proton decoupled $^{13}C$ spectrum of malonic acid in the orientation used in the experiments.  The experiments were performed in a static field of 7.1T using a purpose-built probe.   Also shown is the molecule and the Hamiltonian parameters (all values in kHz).  Elements along the diagonal represent chemical shifts with respect to the transmitter frequency (with the Hamiltonian $\sum_i \pi\omega_i\sigma_z^i$ ).  Elements below the diagonal are dipolar coupling constants ($\sum_{i<j}\frac{\pi}{2}D_{ij}\left(2\sigma_z^i\sigma_z^j - \sigma_x^i\sigma_x^j - \sigma_y^i\sigma_y^j\right)$)  and above the diagonal are J coupling constants ($\sum_{i<j}\frac{\pi}{2}J_{ij}\left(\sigma_z^i\sigma_z^j + \sigma_x^i\sigma_x^j + \sigma_y^i\sigma_y^j\right)$).  The $^{13}C-^{13}C$ values are obtained from the spectral fit.  The relative peak heights give us information about the relative strengths of the dipolar and the indirect J-couplings.  The three central peaks of each multiplet are from the natural abundance of $^{13}C$ present in the molecule at $\approx1\%$.   Combining the fitting information with crystal structure data from neutron scattering experiments \protect\cite{McCalley:1993a} gives the orientation of the molecule with respect to the static magnetic field and from that the proton-carbon dipolar couplings. }
\end{figure}

The majority of previous work in NMR QIP has focussed on liquid state systems that have a simple system Hamiltonian and good coherence properties.  Solid state systems are more difficult to control in practice but offer intrinsically longer coherence times, the ability to pump entropy out of the system of interest into a spin bath and the potential for much higher initial polarizations.   The specific system used here is a three qubit processor molecule, malonic acid \cite{baugh:022305}.  The sample is a macroscopic single crystal, where a small fraction ($\approx 3\%$) of the molecules are triply labeled with $^{13}C$ to form the processor molecules.   The $100\%$ abundant protons in the crystal form the heat-bath.  A proton-decoupled $^{13}C$ spectrum is shown in Fig. \ref{RefSpectrum}.   An accurate natural Hamiltonian is necessary for high fidelity control and is obtained from precise spectral fitting.  The spectrum is simulated from the evolution of the natural Hamiltonian, and the Hamiltonian parameters (chemical shifts, dipolar couplings and the much weaker J couplings, which are usually ignored in the solid-state) are then varied to optimize the fit through a least squares minimization.  The control pulses are designed to be robust to the large uncertainty ($\approx150$Hz) in chemical shift and are fortuitously robust to the much smaller uncertainties ($<10$Hz)  in the coupling results.  

\begin{figure*}[htb]
\includegraphics[scale=0.5]{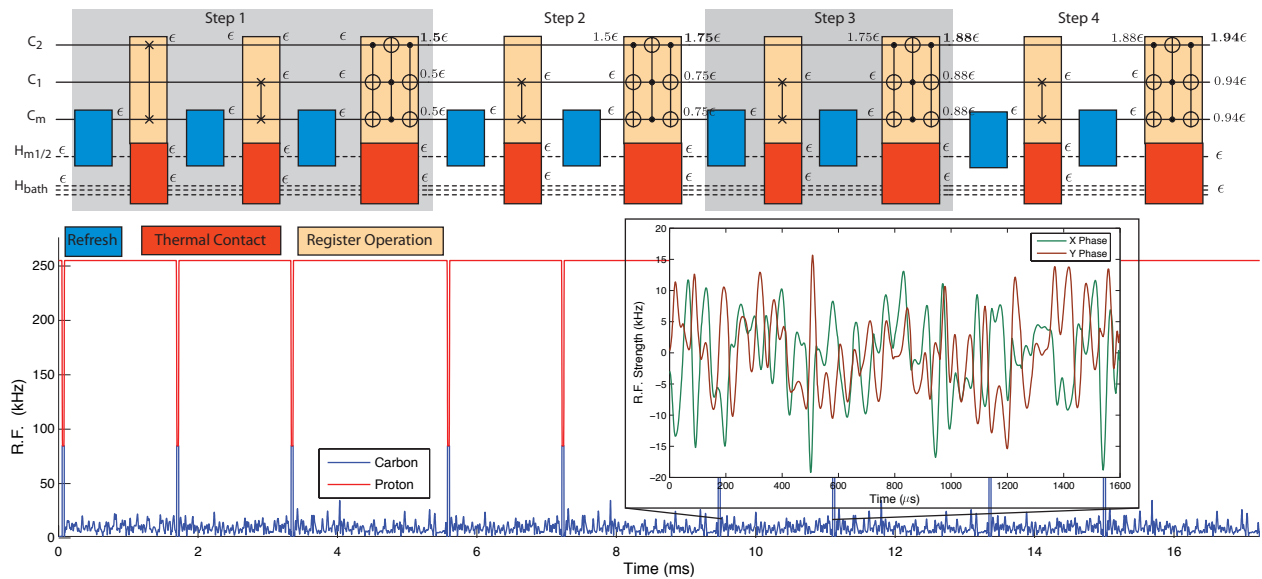}
\caption{\label{Circuit}
The quantum circuit implemented (see text) with the ideal polarizations noted in terms of the heat-bath polarization $\epsilon$.  Each set of swap and compression gate is considered a step and the ideal polarization on the target qubit $C_2$ should increase as 1.5,1.75,1.88,1.94 in steps 1 through 4.    The refresh operations swap polarization from $H_{m1,2}$ to $C_m$ with a short contact CP.  The thermal contact between $H_{m1,2}$ and the rest of the proton bath takes place during the spin locking decoupling pulse for the duration of the carbon register operations.  The swap gates are 1.6ms and the compression gate, 2.2ms.  The compression gate is equivalent to a permutation of the diagonal elements of the density matrix and one possible implementation is shown decomposed as C-NOT-NOT and Toffoli gates \protect\cite{moussa:2005a}; however, it was implemented as a single GRAPE pulse.   The bottom trace shows the amplitude of the radio frequency control fields for the pulse sequence.  The inset shows in detail the two quadrature components of one of the GRAPE control pulses which implements a unitary swap gate between qubits 1 and 3.}  
\end{figure*}

The experiment consisted of four rounds of cooling and compression.  The quantum circuit implemented is shown in Figure \ref{Circuit}.  The carbon register is initialized to infinite temperature by dephasing the thermal polarization.  The bulk $^1H$ polarization was then rotated into the plane and fixed in place with a r.f. spin locking pulse.  Selective transfer of the polarization from $H_{m1,2}$ to $C_m$ served as the refresh step (vide infra).   During the spin locking periods, which also serve to decouple the protons during the carbon-carbon operations, the proton dipolar coupling network still allows for spin diffusion.  Thus, after the refresh $H_{m1,2}$ are cooled by contact with the rest of the proton bath and return to the heat-bath temperature prior to the next refresh step.  The polarization on $C_m$ is swapped onto $C_1$ or $C_2$ with a carbon control sequence.  Once the heat-bath polarization is built up on all three spins, the polarization is then compressed onto $C_2$.  $C_2$ has the smallest proton-carbon coupling and so is least affected by errors during the decoupling and refresh steps and is best able to store the polarization.  Ideally, this first compression step should boost the polarization of $C_2$ to 1.5X the heat-bath polarization and reduce $C_{1/3}$ to 0.5X (end of Step 1 in Fig \ref{Circuit}).   Subsequent steps involve returning $C_m$ and $C_1$ to the heat-bath temperature and repeating the compression step.  In this limit of  the heat-bath polarization $\epsilon \ll 2^{-n}$, the polarization on $C_2$ will asymptotically approach $2\epsilon$ \cite{moussa:2005a}.  

The refresh step is achieved by selectively transferring polarization from the methylene protons $H_{m1,2}$ to the adjacent carbon $C_m$.  Heteronuclear polarization transfer can be achieved through multiple pulse techniques or cross polarization (CP); here we used CP because it was better able to preserve the heat-bath polarization.  Radio frequency fields drive  two spin species at the same nutation frequency (Hartman-Hahn matching condition) which allows them to exchange polarization and their spin temperatures to equalize.   During initial contact the polarization may coherently oscillate between strongly coupled proton and carbon spins \cite{PhysRevLett.32.1402} because, for the relevant input states, the CP condition gives an exchange Hamiltonian.  Thus, if we apply a very short CP pulse, we can selectively swap the polarization from $H_{m1/m2}$ to $C_m$ while negligibly affecting the much more weakly coupled $C_1$ and $C_2$.  Experimentally we found we could increase the polarization on $C_m$ by 3.3X, similar to the enhancement from conventional CP (the theoretical maximum is 3.98X).  When a refresh step was required, the proton spin locking power was smoothly reduced over $10\mu$s to the Hartman Hahn matching condition for $25\mu$s and then smoothly returned to high power.    It should be noted that although CP is the most common method for polarization enhancement of rare spins, it is not the most efficient.  In certain cases,  adiabatic demagnetization may be able to boost the polarization of the rare spins above the heat-bath bias \cite{Lee:2005a}.

The carbon control pulses are optimal control sequences implementing unitary quantum gates even though the PPA requires only classical gates that permute the diagonal elements of the density matrix.  The pulses (see Fig. \ref{Circuit}) are numerically optimized using the GRAPE algorithm \cite{Khaneja:2005fk} - starting from a random guess the pulse is iteratively improved through a gradient ascent search.  In bulk ensembles there are inevitable distributions of control parameters across the sample.  In the current case these cause incoherent loss ($T_2^* \approx 2ms$ ) at a much faster rate the the intrinsic $T_2 \approx 100ms$ \cite{baugh:022305}.    In the present work the most important distributions are the static magnetic field and the r.f. control field.  In order to obtain high experimental fidelities, it was important to demand that the pulses apply the same unitary gate across a range of static fields and pulsing powers.  The GRAPE pulses were numerically optimized to have a fidelity ($|tr(U_{goal}^\dagger U_{sim})|^2/2^{2n}$) of above 0.9975 averaged over a distribution of $\pm 5\%$ in r.f. amplitude and $\pm 150$Hz in static field.  Although the inhomogenities here are specific to ensemble systems, the utility of robust control will be applicable in other single quantum systems for miscalibration, and uncertainty or slow drift in the Hamiltonian.  The pulses were also corrected for non-linearites in the pulse generation and transmission to the sample through the use of a simple feedback circuit which measured the r.f. field at the sample and corrected the pulse accordingly.   The most important element for experimentally achieving high fidelity control was to ensure that the control fields were within the bandwidth of the hardware.  The finite bandwidth of the circuitry produces distortions in both the amplitude and phase of the pulse at switching points \cite{barbara:1991a}.  The solution is to use only smoothly varying control fields.  Although limiting the bandwidth of the optimal control pulses may lead to longer than time-optimal pulses, incoherent sources of decoherence can still be refocussed and higher experimental fidelities result.  

\begin{figure}[htbp]
\includegraphics[scale=0.525]{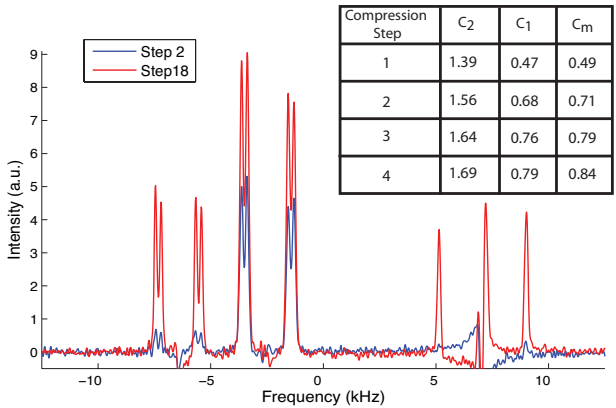}
\caption{\label{Results}
Table of the measured polarization (with respect to the polarization after the initial refresh step) of each spin after the compression gates (steps 6,10,14,18 in the PPA).  All results are $\pm 0.02$.  For the final compression step the heat-bath polarization is no longer needed which allows a switch from the spin locking c.w. decoupling to the more efficient SPINAL64 \protect\cite{Fung:2000lr} (without the switch the enhancement is 1.67X).  The spectra show a comparison of the the first refresh step (swapped to $C_2$) and the final signal after four compression steps.  There is a clear boost of signal on $C_2$, and also substantial polarization on $C_1$ and $C_m$.  The distortions in the spectrum evident for $C_1$ and $C_m$ are due to  residual natural abundance $^{13}$C signal. }
\end{figure}

With these improvements in control we were able to maintain the heat-bath polarization of the proton-spin system and implement repeated rounds of cooling and compression.  The polarization increased for up to four compression steps.  At that point, the polarization of $C_2$ is $1.69\epsilon$ which is well above the Shannon bound of $1.5\epsilon$.  Furthermore, we have built up a non-negligible polarization on the other two qubits of $0.84\epsilon$ and $0.79\epsilon$ increasing the total information content \cite{Brassard:2005a} of the system (see \footnote{See EPAPS Document No. xxxx for figure showing the information content of system through the algorithm.}). 

 Our control is now limited by two factors.  During the carbon control sequences, the protons are decoupled by the spin locking pulse.  This is equivalent to c.w. decoupling which has long been recognized as giving poor decoupling bandwidth as a function of the decoupling power (limited by hardware constraints), particularly if the decoupled spins have strong dipolar couplings as is the case here.  Unfortunately, more efficient decoupling techniques such as SPINAL64 \cite{Fung:2000lr}, do not preserve the magnetization of the decoupled spins which is necessary for this experiment.  We are also limited by the non-ideality of the heat-bath: the proton system is a finite size and so every refresh step heats the bath.  This amount is roughly calculated as the ratio of the number of carbons to protons. Given our 3\% labeling ($^{13}C_3H_4O_4$) this is  $\sim0.5\%$.  Furthermore, there is  relaxation during the spin-locking pulses ($T_1$ in the rotating frame, $T_{1\rho}$) so that our heat-bath is gradually warming during the experiment.   

This experiment represents a step towards creating pure qubits in systems where we have imperfect initialization.  Here we have demonstrated that with sufficient control multiple rounds of cooling and compression can be achieved and the optimal control applied here should be applicable in other QIP systems.  Future work will concentrate on starting with a reset step that has a sufficiently high polarization.  Clearly once errors are considered, perfectly pure qubits are no longer possible.  These experiments, together with our recent work on error characterization \cite{Emerson:2007lr},  suggest an error per gate of  approximately 1\%.  As noted above, this is largely limited by incomplete proton decoupling, a problem specific to this system, and not the control techniques themselves or decoherence.  Even with an error model of a depolarizing rate of 1\% per gate, simulations suggest that with 3/5 qubits, close to pure qubits with polarizations above 97\% are possible with reset polarizations of only 87\%/81\%.  Complete plots of the above threshold scaling behavior are available in Ref. \footnote{See EPAPS Document No. xxxx}.   These polarizations and number of controllable qubits are within reach in a variety of electron-nuclear systems \cite{Negrevergne:2006lr,Hodges:2007a}.  For example, nitrogen vacancy electronic centers in diamond can be optically pumped to $\sim80\%$\cite{Harrison:2006rt}, or the thermal bias of electron spins (g $\approx$ 2) at cryogenic temperatures and typical fields of a few Tesla provides sufficient polarization.

\begin{acknowledgments}
C.R. would like to thank M. Ditty for his technical expertise with the spectrometer.  This work was funded by NSERC, CFI, CIFAR, QuantumWorks, and DTO.
\end{acknowledgments}

\begin{figure}[p]
\includegraphics[scale=0.9]{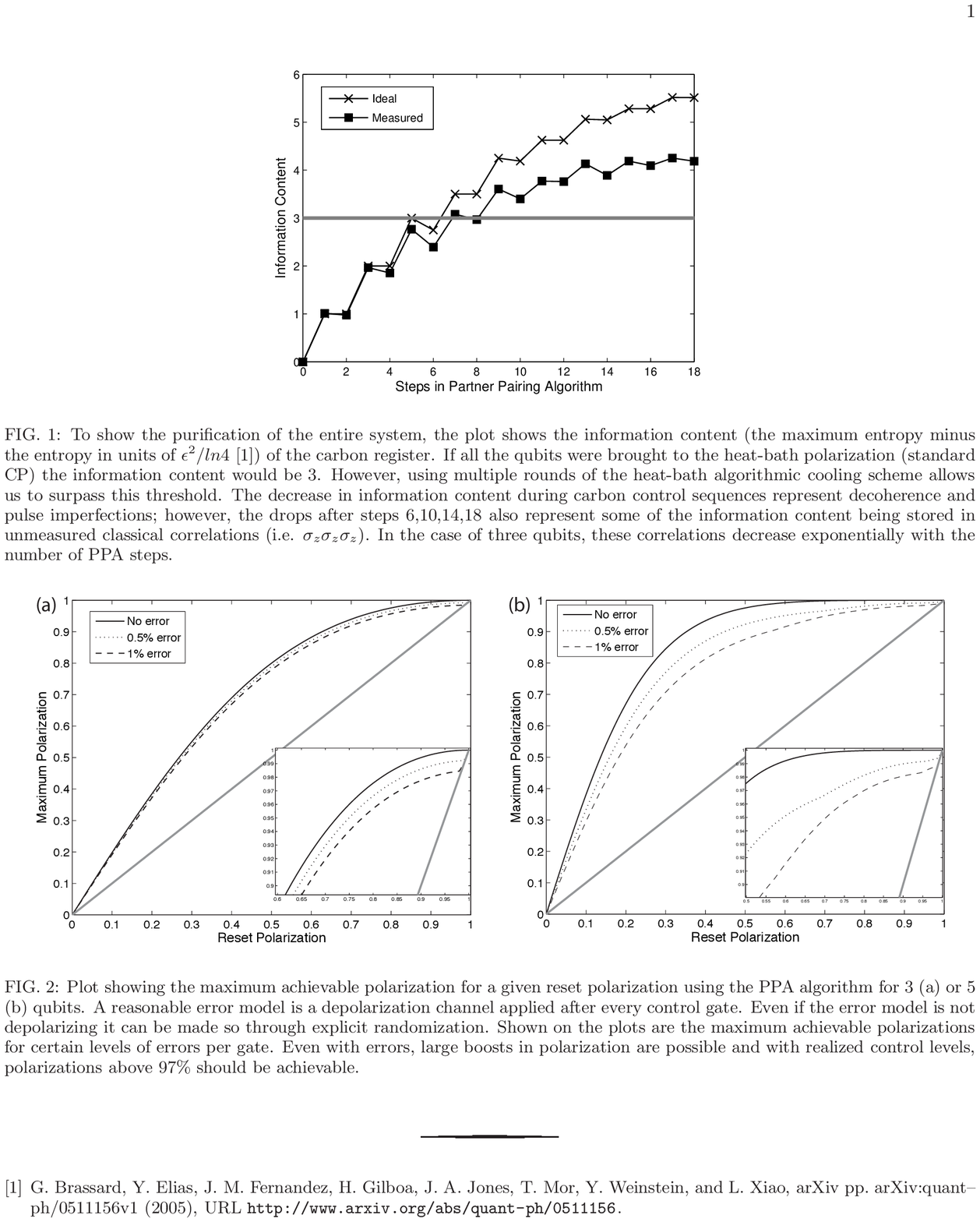}
\end{figure}

\end{document}